\documentclass[preprint,12pt]{elsarticle}

\usepackage{amssymb}
\usepackage{amsmath}
\usepackage{latexsym}
\usepackage{epsfig}
\usepackage{xspace}
\usepackage{color}
\usepackage{epic}
\usepackage{eepic}
\usepackage{algorithmic}
\usepackage{algorithm}
\usepackage{subfigure}

\newtheorem{theorem}{Theorem}

\newtheorem{definition}{Definition}
\newtheorem{example}{Example}

\newenvironment{proof}%
 {\begin{trivlist}\item[]\textbf{Proof.}}%
{\end{trivlist}}

\newcommand{\nop}[1]{}

\newcommand{\stset}{\mathcal{ST}}

\begin{document}
\begin{frontmatter}

\title{Tree Decomposition based Steiner Tree Computation over Large Graphs}

 \author[label1]{Fang Wei-Kleiner}
 \address[label1]{Lin\"oping University, Sweden}

%\numberofauthors{1} 

%\author{
 %1st. author
%\alignauthor
%Fang Wei\\
 %      \affaddr{Computer Science Department}\\
  %   \affaddr{University of Freiburg}\\
    %   \affaddr{Freiburg, Germany}\\
  %    \email{fwei@informatik.uni-freiburg.de}
% 2nd. author
%}

%\maketitle

\begin{abstract}
In this paper, we present an exact algorithm  for the Steiner tree problem.
The algorithm
is based on certain pre-computed index structures.
Our algorithm offers a practical solution
for the Steiner tree problems on graphs of large size and bounded number of terminals.
\end{abstract}

\begin{keyword}
Steiner tree, Graph algorithms, Treewidth, Tree decomposition
\end{keyword}

\end{frontmatter}

\section{Introduction}
%\label{this-chapter-introduction}

%The Steiner tree problem is one of the most prominent  NP-complete
%problems with wide application fields such as network design and keyword search query processing.

The Steiner tree  problem is a well-studied NP-hard problem,
where we have a  graph $G = (V,E)$ with costs on the edges given
and a set of terminals $S \subseteq V$. The goal is to find a minimum-cost tree in $G$ that connects/contains the terminals. 
%The Steiner tree problem is a well-known NP-Hard problem.
% which is among  Karp's original 21 NP-complete problems \cite{Karp72}.
The well-known exact algorithm (parameterized algorithm)  is the 
\emph{Dreyfus-Wagner algorithm} \cite{DW}, which follows the dynamic programming paradigm by computing 
Steiner trees  from its minimum subtrees.
The exact complexity of the algorithm
is $O(|V| \cdot 3^{|S|} +|V|^2 \cdot 2^{|S|} + |V|^3)$.
Hence if $|S|$ is considered as a constant, the algorithm is tractable.

Recently, new applications over Web information systems such as keyword search and social network analysis 
emerge and Steiner tree computation is at the core of the algorithms solving these problems \cite{Li2001WWW}.
One prominent feature in this scenario is that the graph size is large: the size of social networks
or other graph data in the format of XML/RDF  can easily reach hundreds of million of vertices.
As a consequence, for the Web-scale graph data, the parameter $|V|^3$ is dominant and
the computation takes prohibitively long time even $|S|$ is considered as a constant.
Although efforts have been made, algorithms yielding exact results
can only be applied to small size graphs\cite{DPBF}.

%On the other hand, we notice that for these applications, one can build auxiliary
%structures (index structures in database terms)  to accelerate the computation.
%This strategy pays off as long as the index structure  size is linear to the graph and
%the time cost of the Steiner tree computation is improved.

%Recently, the Steiner tree problem has received considerable attention in the database
%community, due to its application in the keyword search query processing
%over graph-structured data.
%These problems can  be naturally defined as a Steiner tree problem, where
%the keywords are considered as the terminals in the graph.
%Although efforts have been made to solve the problem, algorithms yielding exact results
%can only be applied in graphs of small size \cite{DPBF}.
%As a consequence, much research work has been devoted to approximation algorithms,
%such as BANKS~\cite{BANKSI,BANKSII}, BLINKS~\cite{BLINKS}, and STAR~\cite{STAR}.
In this paper, we present an exact algorithm STEIN I  by first constructing 
certain index structures based on the so-called tree decomposition methodology,
and then conducting  the Steiner tree computation 
over the index structure.
We show that our algorithm achieves the run time of
$O(h \cdot (2tw)^{|S|})$ where $tw$ is the treewidth of the graph (see Definition \ref{def:treewidth}),
 and $h$ is the height of the tree decomposition
of $G$ with an upper bound of $|V|$.
%Through the complexity analysis we show that our algorithm offers a practical solution
%for the Steiner tree problems on graphs of large size and bounded number of terminals.
%which is a typical setting in most of the keyword search scenarios.

Chimani et al. \cite{ChimaniMZ12} recently proposed an algorithm for Steiner tree computation
%over graphs with bounded treewidth. 
with the time complexity  
$O(B_{2tw}^2 \cdot tw \cdot |V|)$, where 
$B_{2tw}$ is the \emph{Bell number} with the upper bound of $(2tw)^{2tw}$. 
Clearly this algorithm is only applicable to the graphs with bounded treewidth.
Notice that finding the optimal treewidth of a graph is an intractable problem \cite{Bodlaender93atourist}.
Thus
%for graphs with big size, it is realistic only to obtain certain approximate treewidth values, and 
this algorithm has limitations in practice.
%More related work in this regard can be found in \cite{ChimaniMZ12} as well.

\section{Preliminaries}
\label{sec:pre}
An undirected weighted graph is defined as $G = (V,E)$ plus the weight function $w$, where $V$ is the vertex set and $E \subseteq V \times V$ is the edge set. $w: E \rightarrow \mathbb{Q}^+$ is the weight function. 
%In the rest of the paper, all the graphs are undirected weighted graphs if not otherwise stated.
%We say a vertex $v$ is an element of a graph $G = (V,E)$, denoted as $v \in G$, if and only if $v \in V$ and 
%an edge $e \in G$ if and only if $e \in E$.
%
%Let $G=(V,E)$ be a graph. A path from $v_1$ to $v_n$ in $G$, denoted as $P(v_1,v_n)$,
%consists of a sequence of vertices starting from $v_1$ and ending in $v_n$,
%with the form $v_1, v_2, \ldots, v_n$, where $(v_i,v_{i+1}) \in E$ $(1 \leq i \leq n-1)$.
%Note that all the paths considered in this paper are \emph{simple paths}.
%It is well-known that in a tree, the path between any two vertices is unique.
%Therefore, given two vertices $v_1,v_n$ in a tree $T$, $P(v_1,v_n)$
%is a subgraph $T_P(V_P,E_P)$ where $V_P=\{v_1,v_2,\ldots,v_n\}$
%and $E_P = \{(v_i,v_{i+1}) | 1 \leq i \leq n-1\}$. 
Let $G=(V,E)$ be a graph. $G_1=(V_1,E_1)$ and $G_2=(V_2,E_2)$ be subgraphs
of $G$. The union of $G_1$ and $G_2$, denoted as $G_1 \cup G_2$,
is the graph $G'=(V',E')$ where $V' = V_1 \cup V_2$ and $E' = E_1 \cup E_2$.

%To simplify the representation, given a vertex $v$ and an edge $e$,
%we write
%$v \in G$ to mean that $v$ is a vertex of $G$,
%and  $e \in G$ to mean that $e$ is an edge of $G$.
%In the following we introduce the definition of subtree by a set of vertices. 

\begin{definition} [Steiner tree]
Let  $G = (V,E)$ be an undirected graph with the weight function $w$. $S \subseteq V$ 
is a set of terminals. The Steiner tree with respect to $S$,
denoted as $ST_S$, is a tree 
spanning the terminals, where $w(ST_S) := \sum_{e \in ST_S}w(e)$ is minimal.
\end{definition}

\begin{figure*}[ht]
 \centering
 \subfigure[Graph]{
\includegraphics [scale=0.45]{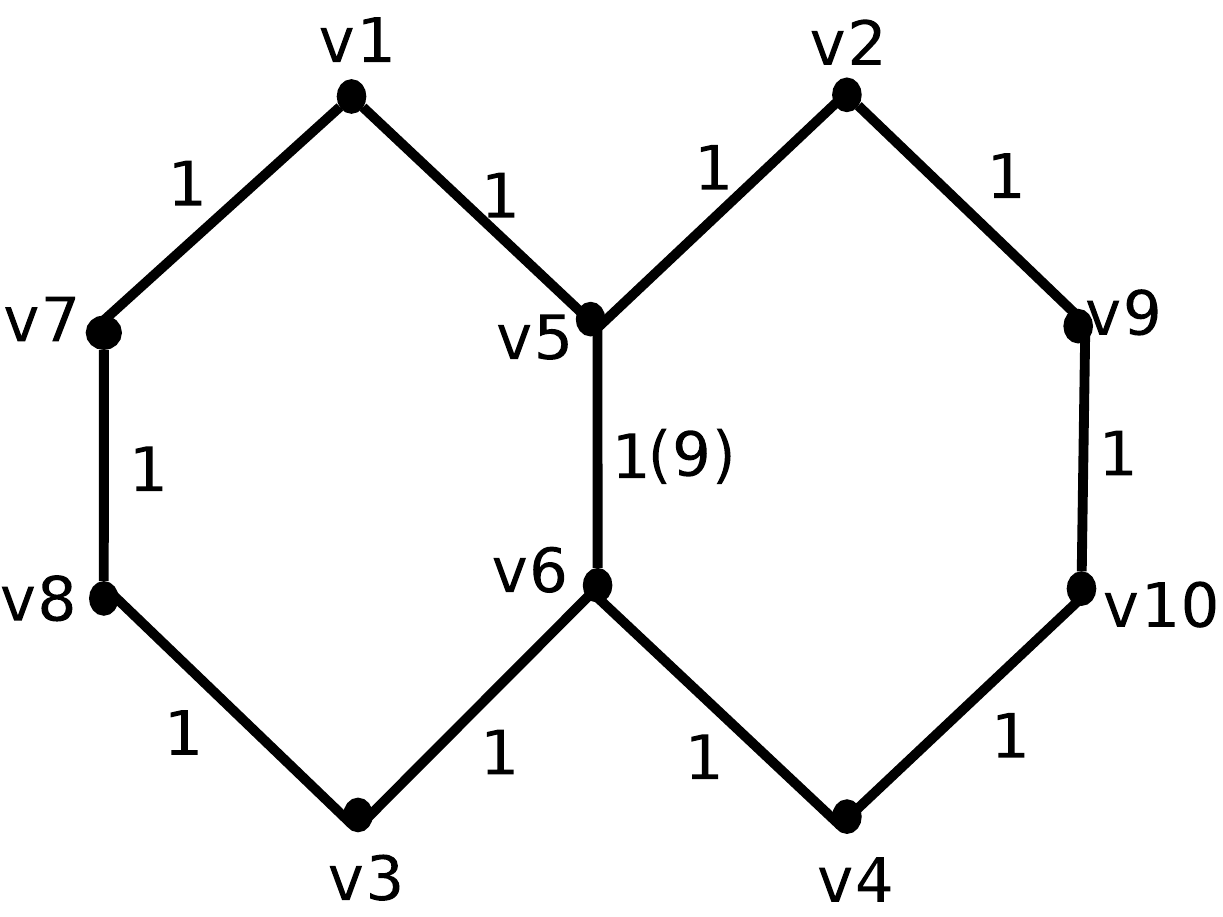}
   \label{fig:subfig1}
   }
\subfigure[Tree decomposition]{
  \includegraphics [scale=0.26]{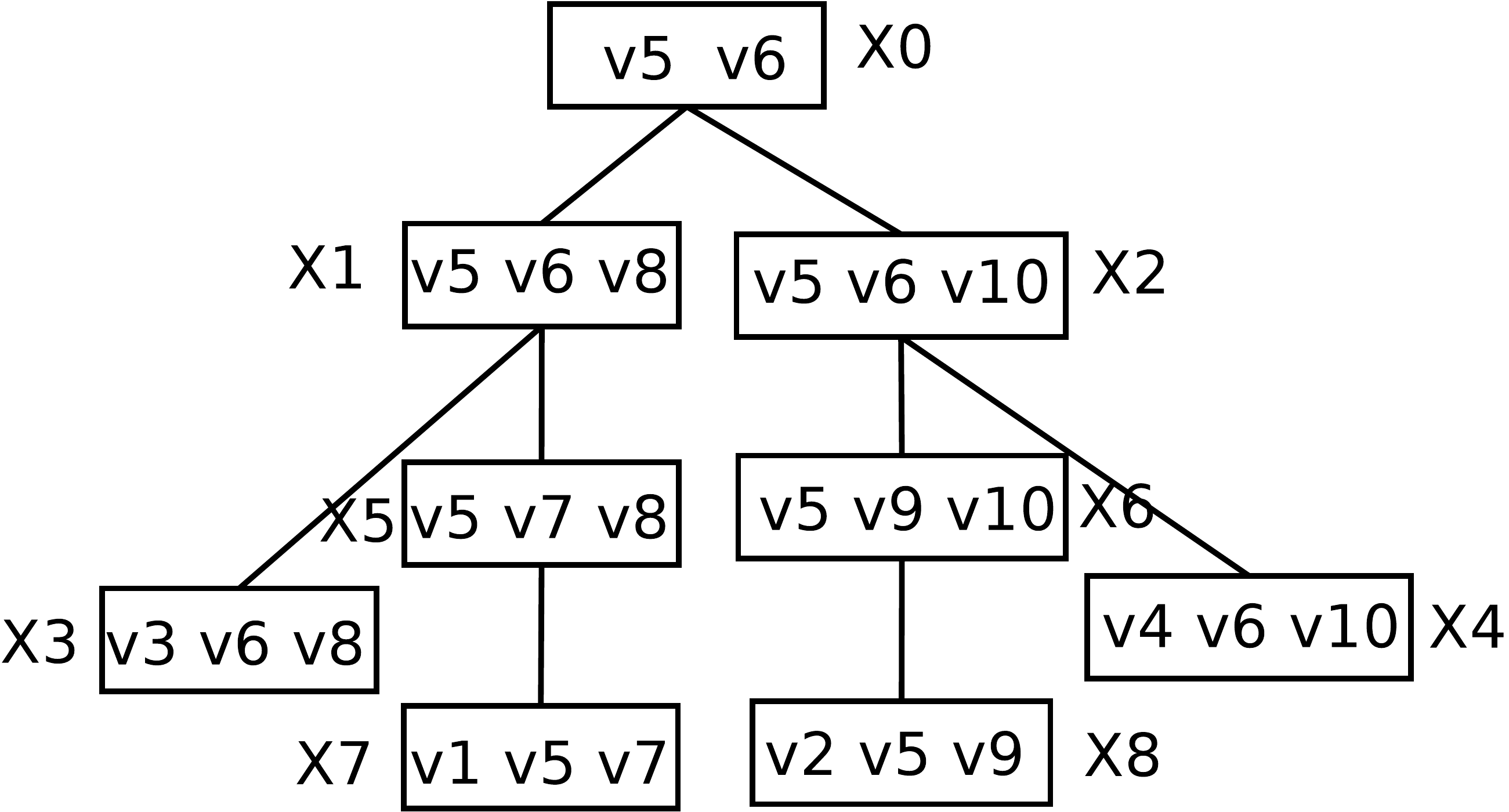}
  \label{fig:decomposition}
}
\label{fig:subfigureExample}
\caption{Example graphs with $S=\{v_1, v_2, v_3, v_4\}$ as terminals and the tree decomposition.
%\subref{fig:subfig1} Graph distinguishes from \subref{fig:subfig2} Graph  at the edge weight of $(v_5,v_6)$.
%        Steiner tree for \subref{fig:subfig1} Graph contains the edge $e(v_5,v_6)$ with $w(ST_S)$ as 5,   while Steiner tree for \subref{fig:subfig2} Graph  is the path of
  %      $(v_2, v_5, v_1, v_7, v_8, v_3, v_6, v_4)$ with $w(ST_S)$ as 7.
  }
 \label{fig:running}      
\end{figure*}

%The difference between the Steiner tree problem and the minimum spanning tree problem is that in the Steiner tree problem extra intermediate vertices and edges may be added to the graph in order to reduce the length of the spanning tree. These new vertices introduced are called as \emph{Steiner vertices}.
If the context is clear, we will  sometimes use the statement  "$ST_S$ has the value of" by meaning that "the weight of $ST_S$ has the value of".
As a running example, consider the graphs illustrated in  Figure \ref{fig:subfig1},
where two graphs are illustrated in the same figure and they distinguish from each other on the 
weight of the edge $(v_5,v_6)$, where graph 1 has the weight 1 and graph 2 has the value  9.
Assume  $S=\{v_1, v_2, v_3, v_4\}$.
Steiner tree for Graph 1 has the weight 5 including $(v_5,v_6)$ while the Steiner tree for Graph 2 does not include $(v_5,v_6)$.
%Steiner tree  $ST_S$ for Graph 1 consists of the 
%edges $(v_1,v_5)$, $(v_2,v_5)$, $(v_5,v_6)$, $(v_3,v_6)$ and $(v_4,v_6)$,
%with the weight of 5. On the other hand, the Steiner tree in Graph 2 does not contain
 %the edge $(v_5,v_6)$. One possible Steiner
%tree is  the path 
%$(v_2$, $v_5$, $v_1$, $v_7$, $v_8$, $v_3$, $v_6$, $v_4)$ with $w(ST_S)$ as 7.

\subsection{Algorithm STVS}
\label{sec:stvs}

In this section, we introduce the first Steiner tree algorithm STVS.

\begin{definition} [Vertex Separator]
Let $G=(V,E)$ be a graph, $v_0, v \in V$. $C \subseteq V$ is a 
$(v_0, v)$-vertex separator, denoted as $(v_0, v)$-VS, if for every path $P$ from $v$ to $v_0$,
there exists a vertex $u$ such that $u \in P$ and $u \in C$.
\end{definition}

\begin{theorem}
Let $G=(V, E)$ be a graph, $v, v_0 \in V$, $S \subseteq V$ and $S = \{v_1, \ldots, v_n\}$.
$C \subseteq V$ is a $(v, v_0)$-VS. Then 
\begin{equation}
\label{equ:stvs1}
ST_{S \cup v_0 \cup v} = {\mbox{min}} ~~~ ST_{S' \cup w \cup v}  \cup ST_{S'' \cup w \cup v_0}
\end{equation}
where minimum is taken over all $w \in C$ and all  bipartitions
$S = S' \cup S''$.
\label{the:stvs}
\end{theorem}

\begin{figure}[ht]
 \centering
\subfigure[Illustration of Theorem \ref{the:stvs}]{
  \includegraphics [scale=0.36]{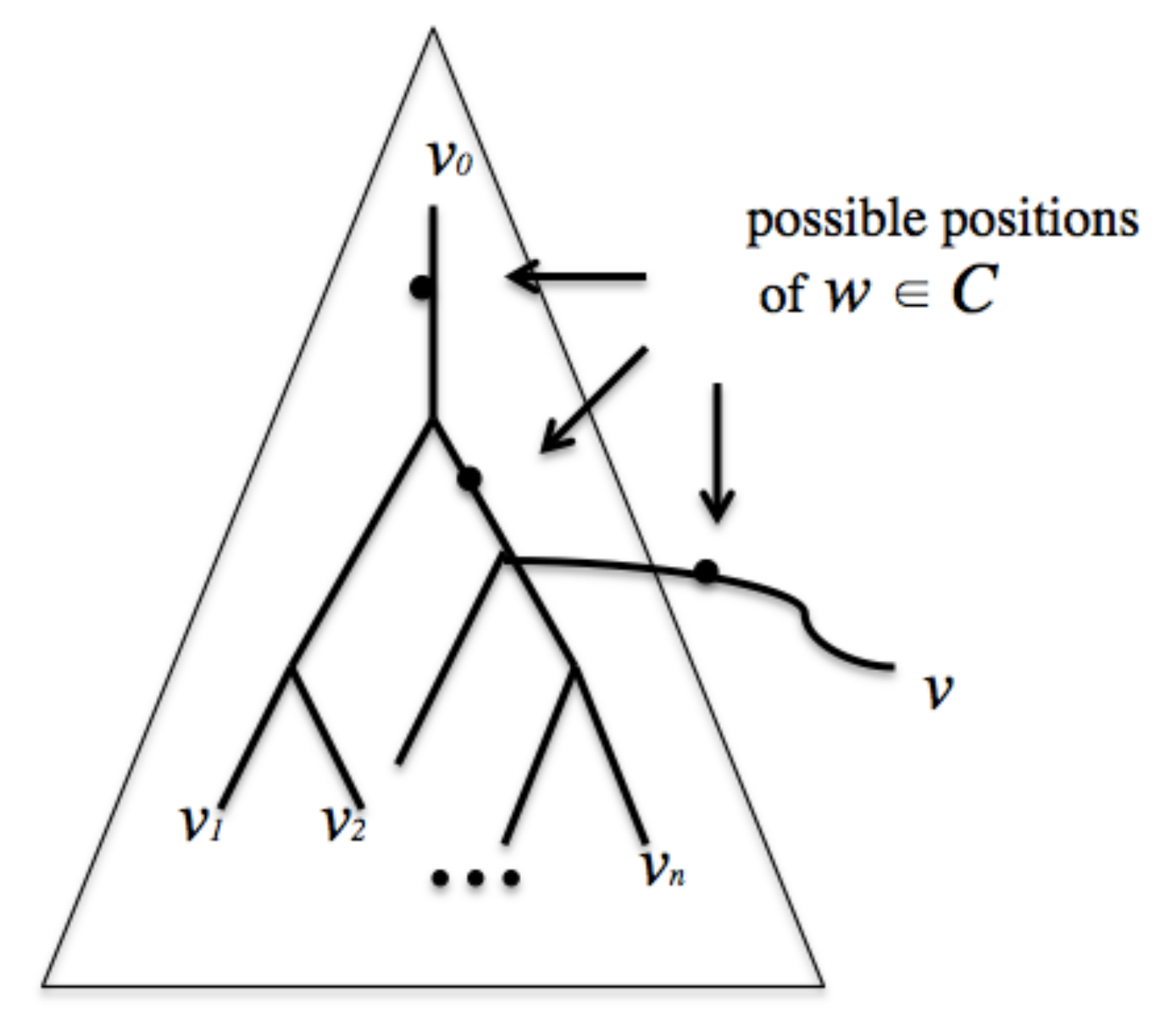}
  \label{fig:stein1}
}
\subfigure[Nice tree decomposition]{
	\includegraphics [scale=0.20]{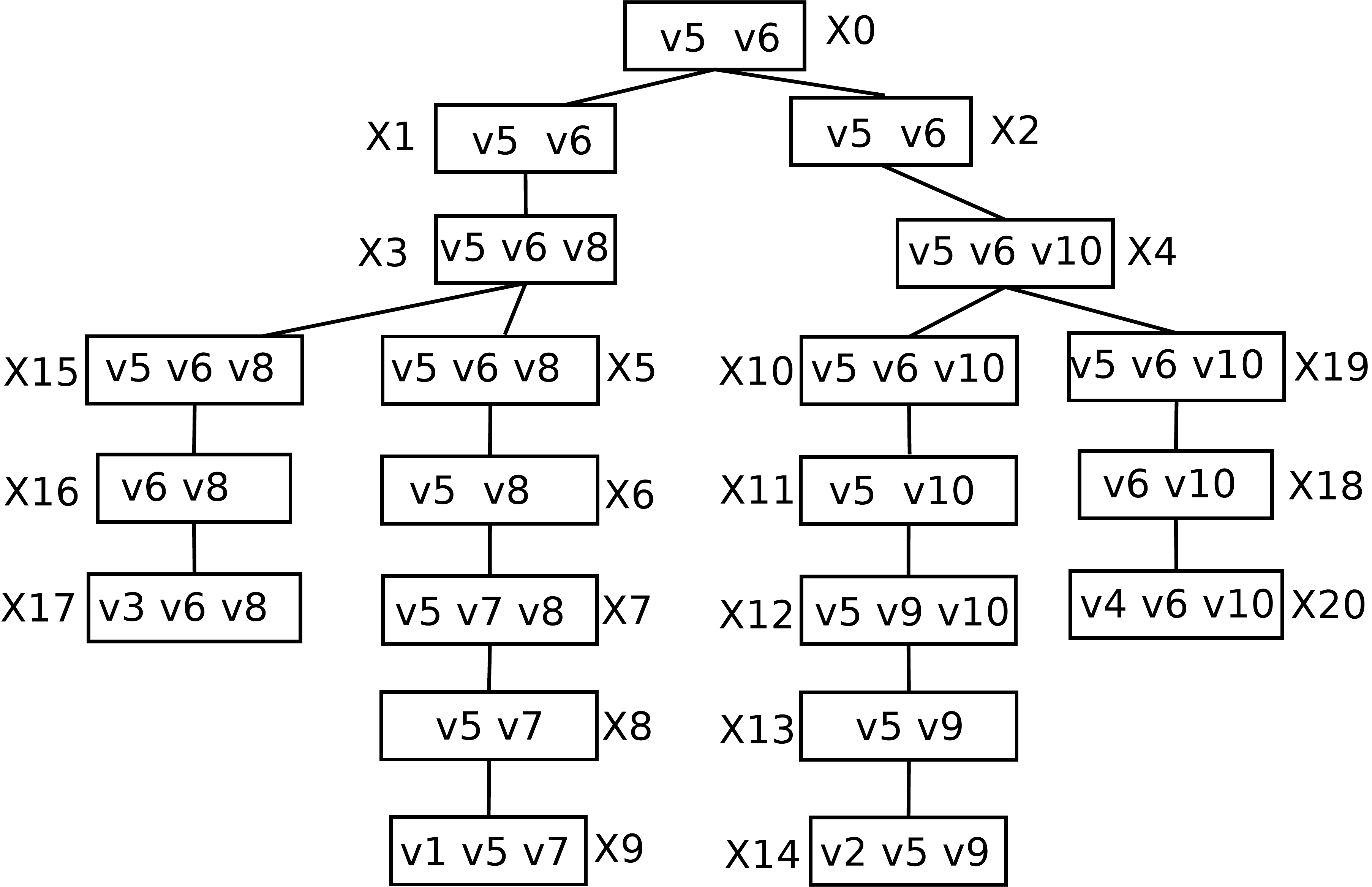}
%	\caption{Nice tree decomposition of example graph}
	\label{fig:nicedecomposition}
}
\end{figure}

%\begin{figure}[ht]
%	\centering 
%	\includegraphics [scale=0.35]{pic/stein1}
%	\caption{Illustration of Theorem \ref{the:stein1}}
%	\label{fig:stein1}
%\end{figure}

\begin{proof}
Consider the Steiner tree $ST_{S \cup v_0 \cup v}$.
There must exist a path $P$ from $v$ to $v_0$.
Given the fact that $C \subseteq V$ is a $(v, v_0)$-VS, 
we know that there exists one vertex $w \in C$, such that $w \in P$ as shown in Figure \ref{fig:stein1}.
No matter where $w$ is located, we can split $ST_{S \cup v_0 \cup v}$
into two subtrees. One is the subtree rooted at $w$, which contains $v$.
The other subtree contains
the rest of the terminals in $S$, together with $v_0$ and $w$. 
Each of the subtree is a Steiner tree regarding the terminals. 
It is trivial to show the minimum of both trees, due to the fact that
the entire tree is a Steiner tree. \qed
\end{proof}

\begin{algorithm}
\begin{algorithmic}
\REQUIRE{$G=(V,E)$, $v, v_0 \in V$, $S = \{v_1, \ldots, v_n\}$,
$C \subseteq V$ is a $(v, v_0)$-VS. }
\ENSURE{$ST_{S \cup v_0 \cup v}$}

\bf{for all} {vertex $w$ $\in$ $C$}

~~\bf{for all} {$S'$, $S''$ where  $S = S' \cup S''$}

~~~~~~ {$ST_{S \cup v_0 \cup v} = {\mbox{min}} ~~~ ST_{S' \cup w \cup v}  \cup ST_{S'' \cup w \cup v_0}$}
%\ENDFOR
%\ENDFOR
\RETURN {$ST_{S \cup v_0 \cup v}$}
\end{algorithmic}
\caption{STVS($v, v_0, S, C$)}
\label{alg:stvs}
\end{algorithm}

Algorithm \ref{alg:stvs} shows the pseudo code for computing the Steiner tree according 
to Theorem \ref{the:stein1}. The complexity of the algorithm is $|C| \cdot 2^{|S|}$.
One important observation about STVS   is that
the number of terminals of the sub-Steiner trees is not necessarily less than
that of the final Steiner tree $ST_{S \cup v_0 \cup v}$. 
For instance, take the case of $S' = S$ and $S'' = \emptyset$,
the number of terminals of $ST_{S' \cup w \cup v}$ is equal to 
$ST_{S \cup v_0 \cup v}$ (both are $|S|+2$). Therefore, the dynamic programming
paradigm is not applicable in this regard.
Moreover, given only the graph, it is unknown how to compute the vertex separator set $C$.
If $C$ is not confined in any form (i.e. $C=V$), then  STVS becomes
an algorithm \'a la Dreyfus-Wagner.
Therefore, in order to make the STVS algorithm useful, it has to be guaranteed that all 
the sub-Steiner trees be pre-computed, and $C$ be relatively small comparing to $V$.
In the following section, we will explain in detail how these conditions 
are fulfilled with the tree decomposition techniques. 

\subsection{Tree Decomposition and Treewidth}
%In this section, we first introduce the basic definitions of tree decomposition,
%then we propose the algorithm over the tree decomposition based on 
%Theorem \ref{the:stein1}.

\begin{definition} [Tree Decomposition]
A tree decomposition  of a graph $G = (V, E)$, denoted as $T_G$, is a pair $(\{X_i | i \in I\}, T)$, where $I$ is a finite set of integers with the form $\{0, 1, \ldots, p\}$ and $\{X_i | i \in I\}$ is a collection of subsets of $V$ and $T = (I,F)$ is a tree such that:
\begin{enumerate}
	\item  $\bigcup_{i \in I} X_i = V$.
	\item for every $(u, v) \in E$, there is $i \in I$, s.t. $u, v \in X_i$.
	\item for every $v \in V$, the set  $\{i \vert$ v $\in X_i\}$ forms a connected subtree of  $T$.
\end{enumerate}
\label{def:TD}
\end{definition}

A tree decomposition consists of a set of tree nodes, where each node contains a set of vertices in $V$. We call the sets $X_i$ \emph {bags}. It is required that every vertex in $V$ should occur in at least one bag (condition 1), and for every edge in $E$, both vertices of the edge should occur together in at least one bag (condition 2). The third condition is usually referred to as the connectedness condition, which requires that given a vertex $v$ in the graph, all the bags which contain $v$ should be connected.

Note that from now on, the node in the graph $G$ is referred to as vertex, and the node in the tree decomposition is referred to as tree node or simply node. For each tree node $i$, there is a bag $X_i$ consisting of vertices. To simplify the representation, we will sometimes use the term node and its corresponding bag interchangeably. 
%Given a tree decomposition $T_G$, we denote its root as $R$.

%\begin{figure}[ht]
%	\centering 
%	\includegraphics [scale=0.25]{pic/td}
%	\caption{Example graph and tree decomposition}
%	\label{fig:decomposition}
%\end{figure}

Figure \ref{fig:decomposition} illustrates a tree decomposition of the graph from the running example.
% Since the two graphs in Figure \ref{fig:running} distinguish from each other only on the weight of one edge, they have the identical tree decomposition (the weight in the graph is irrelevant to the tree decomposition).
In most of the tree decomposition related literature, the so-called \emph{nice tree decomposition}
 is used. In short, a nice tree decomposition is a tree decomposition, with 
 the following additional conditions:
(1) Every internal node $t \in T$ has either 1 or 2 child nodes.
(2) If a node $t$ has one child node $t_0$, then the bag $X_t$ is obtained from $X_{t_0}$ either
by removing one element or by introducing a new element. (3) If a node $t$ has two
child nodes then these child nodes have identical bags as $t$.
Given a tree decomposition $T_G$,  the size of the  nice tree decomposition of 
$T_G$ is linear to it. Moreover, the transformation can be done in linear time
w.r.t. the size of $T_G$. 
%See \cite{Bodlaender93atourist} for a detailed introduction for tree decomposition based algorithms.
%The purpose of introducing nice tree decomposition is to simplify the bag
%traversal algorithm. As mentioned above, tree decomposition based algorithm commonly
%follow the the traversal procedure in a bottom up manner. As a consequence,
%if we traverse over a nice tree decomposition, each time from the child bag to
%the parent bag involves simply one vertex inserted, or one vertex removed.
Figure \ref{fig:nicedecomposition} shows the nice tree decomposition of the running example graph. 

%\begin{figure}[ht]
%	\centering 
%	\includegraphics [scale=0.23]{pic/nicetd}
%	\caption{Nice tree decomposition of example graph}
%	\label{fig:nicedecomposition}
%\end{figure}

%It consists of bags $X_0, \ldots, X_8$, and all these bags are connected in a form 
%of a tree. We assume $X_0$ as the root of the tree. In each bag, there is a set of vertices
%from $V$. Note that three conditions in Definition \ref{def:TD} are fulfilled in this example.
%One interesting observation is that a given vertex can occur in multiple bags.
%Intuitively, the more neighbors a vertex has in $G$, the more bags it occurs in $T_G$. Because
%we need to reflect every  edge information in the bags.
%On the other hand,  the connectedness property of the tree decomposition
%(condition 3 in Definition \ref{def:TD})
%enforces the \emph{locality} of the vertices in the bags.
%That is,  
%every vertex $v$ in $G$ has its \emph{induced subtree} in $T_G$,
%in which every bag contains $v$.  In another word,
%every vertex can be found only in the bags of the induced subtree in $T_G$,
%and nowhere else.

\begin{definition} [Induced Subtree]
Let $G=(V,E)$ be a graph and $T_G$ its tree decomposition. $v \in V$. 
The induced subtree of $v$ on $T_G$,
denoted as $T_v$, is a subtree of $T_G$ such that
for every bag $X \in T_G$,  
$v \in X$ if and only if $X \in T_v$.
\end{definition}

Intuitively, the induced subtree of a given vertex $v$ consists of precisely those bags that contain $v$.
Due to the connectedness condition, $T_v$ is a tree.
With the definition of induced subtree, any vertex $v$ in the graph $G$ can be uniquely
identified with the root of its induced subtree in $T_G$.
Therefore, from now on we will use the expression of "the vertex $v$ in $T_G$" with the intended meaning
that "the root of the induced subtree of $v$ in $T_G$", if the context is clear.

The following theorem reveals the the relationship between a tree decomposition structure 
and the vertex separator.

\begin{theorem} \cite{TEDI}
Let $G=(V,E)$ be a graph and $T_G$ its tree decomposition. 
$u, v \in V$.
%Let $T_u$ and $T_v$ be the induced subtree of $u$ and $v$ in $T_G$,
%and $r_u$ and $r_v$ be the root of $T_u$ and $T_v$ respectively.
%Let $X$ be a bag from the tree path between $r_u$ and $r_v$ on $T_G$.
Every bag $X$ on the path between $u$ and $v$ in $T_G$  
is a $(u,v)$-vertex separator.
\label{the:vsintd}
\end{theorem}

%For instance, the induced subtree of $v_5$ contains the bags 
%$X_0$, $X_1$, $X_2$, $X_5$, $X_6$, $X_7$ and $X_8$,
%whereas the induced subtree of $v_9$ contains $X_6$ and $X_8$,
%with $X_6$ as the root. 

%The connectedness condition of the tree decomposition reveals the most important
 %design philosophy for algorithms based on the tree decomposition.
 %Namely any single vertex is located in a connected set of bags in 
 %the tree decomposition. In general, tree decomposition based 
 %algorithm follows the bottom up processing on bags. Thus as long as 
 %the root of the induced subtree of a vertex $v$ is processed, $v$ will be
 %\emph{removed} from the current data structure. Because it will never occur
 %in the rest of the tree.
%Therefore, if the bag size is bounded, the search space for 
%the algorithm is restricted.
%In fact,
%many intractable graph problems can be solved in polynomial or even linear time,
%as long as the treewidth (definition see below) of the underlying graph is bounded, which means,
%can be considered as a constant. 

%Given any graph $G$, there may exist many tree decompositions which fulfill all the conditions in Definition \ref{defiTD}. However, we are interested in those tree decompositions with smaller bag sizes. We call the cardinality of a bag the \emph{width} of the bag.

\begin{definition} [Width, Treewidth] 
Let $G = (V,E)$ be a graph.
The \emph{width} of a tree decomposition $(\{X_i \vert i \in I\}, T)$ is defined as $max\{\vert X_i \vert -1 \  \vert i \in I\}$.
The \emph{treewidth} of $G$ is the minimal width of all tree decompositions of $G$. It is denoted as $tw(G)$ or simply $tw$.
\label{def:treewidth}
\end{definition}

%Again consider our example: the width of the tree decomposition is 3.
%Since only trees have the treewidth of 2, and obviously our graph in
%Figure \ref{fig:running} is not a tree, we can conclude that the width of 3 is optimal,
%thus the treewidth of the graph is 3.

%\noindent {\bf Nice tree decomposition}

\section{STEIN I}

\nop{
\begin{example}
Consider the nice tree decomposition in Figure \ref{fig:nicedecomposition}.
The root of the induced subtree of vertices $v_3$ and $v_{10}$
is $X_{17}$ and $X_4$ respectively.
Then according to Theorem \ref{the:vsintd}, every one from the bags along the path
between $X_{17}$ and $X_4$, namely $X_{17}$, $X_{16}$, $X_{15}$,
$X_{3}$, $X_1$, $X_0$, $X_2$ and $X_4$, is a $(v_3,v_{10})$-vertex separator.
\end{example}
}
%Intuitively, Theorem \ref{the:vsintd} restricts the search space between any two 
%vertices $u$ and $v$ in the graph $G$, in the sense that \emph{every} path
%between $u$ and $v$ must pass through all the bags along the path from $r_u$ to $r_v$.

\begin{definition}[Steiner Tree Set]
Given a set of vertices $S=\{v_1,\ldots,v_n\}$,
the Steiner tree Set $\stset_S^m$ is the set of the Steiner trees
of the form $ST_{u_1,\ldots,u_k}$ where $\{u_1,\ldots,u_k\}$ $\subseteq$ $\{v_1,\ldots,v_n\}$
and $2 \leq k \leq m$.
\end{definition}

Now we  are ready to present the algorithm STEIN I,
which consists of mainly two parts:
(1) Index construction, and  (2) Steiner tree query processing.
In step (1), we first generate the tree decomposition $T_G$ for a given graph $G$.
Then for each bag $X$ on $T_G$,
we compute   $\stset_X^l$, where $l$ is the number 
of terminals of the Steiner tree computation.
In another word, for computing a Steiner tree with $l$ terminals, 
we need to pre-compute in each bag all the Steiner trees with 2, 3,.$\ldots$, $l$ terminals.

\begin{theorem}[STEIN I]
\label{the:stein1}
Let $G=(V,E)$ and $T_G$ is the tree decomposition of $G$ with treewidth $tw$.
$S \subseteq V$ is the terminal set. For every bag $X$ in $T_G$,
$\stset_X^{|S|}$ is pre-computed. Then $ST_S$ can be computed
in time $O(h \cdot (2tw)^{|S|})$, where $h$ is the height of $T_G$.
\end{theorem}

%Theorem \ref{the:stein1} states that (1) the Steiner tree can be computed
%given that the sub-Steiner trees in the bags of $T_G$ are available, and
%(2) the time complexity is linear to $h$, if we consider both treewidth
%of the graph and the number of terminals as constants.
%In the following, we will prove Theorem \ref{the:stein1} by induction.

\begin{proof}
Assume that $T_G$ is a nice tree decomposition.
First, for each terminal $v_i$ we  identify the root of the induced subtree $X_i$ in $T_G$.
Then we retrieve the lowest common ancestor (LCA) of all $X_i$.
We start from the $X_i$s, conduct the bottom up traversal from the children nodes to the parent node
over $T_G$, till LCA is reached.

Given a bag $X$ in $T_G$, we denote all the terminals located in the subtree rooted at $X$ as $S_X$.
In the following we prove the theorem by induction.

\noindent
{\bf Claim:} Given a bag $X$ in $T_G$, if for all its child bags $X_i$,
$\stset_{X_i \cup S_{X_i}}^{|S|}$ are computed, then 
$\stset_{X \cup S_{X}}^{|S|}$ can be computed with the time $O( (2tw)^{|S|})$.

\noindent
{\bf Basis:} Bag $X$ is the root of the induced subtree of a terminal $v$, and there is no other terminal below $X$.
(That is, $X$ is the one of the bags where we start with.) 
In this case, $X \cup S_{X}$ = $X$ and $\stset_{X \cup S_{X}}^{|S|}$ = $\stset_{X}^{|S|}$.
This is exactly what was pre-computed in bag $X$.

\noindent
{\bf Induction:} 
%Assume that given a bag $X$ in $T_G$, 
%$\stset_{X_i \cup S_{X_i}}^{|S|}$ are computed for all its child bags $X_i$,
%we need to prove that $\stset_{X \cup S_{X}}^{|S|}$ can be computed with the time complexity
%$O( (2tw)^{|S|})$.
In a nice tree decomposition, there are three traversal patterns from the child nodes
to the parent node:
%\begin{itemize}
%\item 

\noindent (*) Vertex removal:
parent node $X$ has one child node $X_c$ where $X_c = X \cup v$.
%\item 

\noindent (*) Vertex insertion: 
parent node $X$ has one child node $X_c$ where $X = X_c \cup v$.

%\item 

\noindent (*) Merge: 
parent node $X$ has two child nodes $X_{c_1}$ and $X_{c_2}$,
where $X$= $X_{c_1}$ = $X_{c_2}$
%\end{itemize}
%In the following we will prove the claim regarding the above three patterns.
%For a better understanding, we take the nice tree decomposition 
%from Figure \ref{fig:nicedecomposition} as our running example 
%and assume the terminals are $v_1$, $v_2$,
%$v_3$ and $v_4$. 

\noindent
{\bf Vertex removal}. Assume the current parent bag $X$ has the child bag
$X_c$, such that $X_c = X \cup v$ .
%(see e.g. the traversal from $X_9$ to $X_8$ in  Figure \ref{fig:nicedecomposition}). 
We observe that $S_{X}$ = $S_{X_c}$. That is, the terminal set below $X$ 
remains the same as with $X_c$. This is because from $X_c$ to $X$, no new vertex
is introduced. There are two cases:
\begin{itemize}
\item
$v$ is a terminal. Then we need to remember $v$ in $X$.
%because it is a terminal below $X$. 
Therefore, we have $\stset_{X \cup S_{X}}^{|S|}$ = 
$\stset_{X_c \setminus v  \cup S_{X} \cup v}^{|S|}$ =
$\stset_{X_c \cup S_{X_c}}^{|S|}$.
That is, the Steiner tree set in $X$ remains exactly the same as $X_c$.
%(e.g. the traversal from $X_9$ to $X_8$ in  Figure \ref{fig:nicedecomposition} belongs to this case).
\item
$v$ is not a terminal. 
%For instance, the traversal from $X_7$ to $X_6$ in 
%Figure \ref{fig:nicedecomposition}, where $v_7$ is removed and $v_7$ 
%is not a terminal. 
In this case, we remove simply all the Steiner trees
from $\stset_{X_c \cup S_{X_c}}^{|S|}$ where $v$ occurrs as a terminal.
This operation costs constant time.
\end{itemize}

\noindent
{\bf Vertex insertion}. Assume the current parent bag $X$ has the child bag
$X_c$, such that $X = X_c \cup v$. 
%(see e.g. the traversal from $X_8$ to $X_7$ in  Figure \ref{fig:nicedecomposition}). 
%This is a more complex case and Algorithm \ref{alg:stvs} is applied.
First let us consider the inserted vertex $v$ in $X$. Note that $v$ does not occur
in $X_c$, so according to the connectedness condition, $v$ does not occur
in any bag below $X_c$. Now consider any terminal  $v_i$ below $X_c$. 
According to the definition, the root of the induced subtree of $v_i$ is also below $X_c$.
In the following we first prove that $X_c$ is a $(v, v_i)$-vertex separator.
%(See e.g. $X_8$ is a $(v_1,v_8)$-vertex separator in Figure \ref{fig:nicedecomposition}).

As stated above, $v$ occurs in $X$ and does not occur in $X_c$, so
we can conclude that the root of the induced subtree of $v$  ($r_v$) is an ancestor 
of $X$. Moreover, we know that the root of the induced subtree of $x_i$  ($r_{v_i}$) is below
$X_c$. As a result, $X_c$ is in the path between $r_v$ and $r_{v_i}$.
Then according to Theorem \ref{the:vsintd}, $X_c$ is a $(v, v_i)$-vertex separator.

Next we generate all the Steiner trees $ST_{Y \cup v}$, where
$Y \subseteq X_c \cup S_{X_c}$ . 
We execute the generation of all the Steiner trees in an incremental manner, 
by inserting the vertices in $X_c \cup S_{X_c}$ one by one to $v$, starting from
the vertices in $X_c$, which is then followed by $S_{X_c}$.
%For instance, for the traversal from $X_8$ to $X_7$, we generate the Steiner 
%trees in the following order:
%$\stset_{v_5,v_8}^4$, $\stset_{v_5, v_7, v_8}^4$, $\stset_{v_5, v_7, v_1, v_8}^4$.
We distinguish the following cases:
\begin{itemize}
\item
$Y \subseteq X_c$. That is, all vertices in $Y$ occurs in $X_c$. Obviously 
$ Y \subseteq X$ holds. Then the Steiner tree $ST_{Y \cup v}$ is pre-computed
for bag $X$ and we can directly retrieve it.
\item 
$Y \cap S_{X_c} \neq \emptyset$. According to the order of the Steiner trees generation above, we assign the newly inserted terminal as $v_i$. 
Let $W = Y \setminus v_i$.
It is known that $\stset_{W \cup v}^{|S|}$ is already generated.
We call the function STVS as follows: 
STVS$(v_i, v, W, X_c)$. Since $X_c$ is the $(v_i,v)$-vertex separator as we shown above,
the function will correctly compute the results, as long as all the sub-Steiner trees are available.
Let $W=W' \cup W''$. The first sub-Steiner tree $ST_{W' \cup w \cup v_i}$ (where $w \in X_c$)
can be retrieved from the Steiner tree set in $X_c$, because there is no $v$ involved.
The second sub-Steiner tree has the form $ST_{W'' \cup w \cup v}$ (where $w \in X_c$).
It can be retrieved from $\stset_{W \cup v}^{|S|}$, because $W'' \subseteq W$ and 
$X_c \subseteq W$. The later is true due to the fact that the current step by inserting terminals,
all the vertices in $X_c$ have already been inserted, according the order of Steiner tree generation we assign.
This complete the proof of the correctness.
\end{itemize}

As far as the complexity is concerned, at each step of vertex insertion, 
we need to generate $tw^{|S|-1}$ new Steiner trees. Each call of the function of STVS
takes time $tw \cdot 2^{|S|}$ in worst case. Thus the total time cost 
is $tw^{|S|} \cdot 2^{|S|}$ = $(2tw)^{|S|}$.

\noindent 
{\bf Merge}: Merge operation occurs as bag $X$ has two child nodes $X_{c_1}$ and $X_{c_2}$,
and both children consist of the same set of vertices as $X$.
Since each of the child nodes induces a distinct subtree, the terminals below the child nodes
are disjunctive. 
%For example, bag $X_3$ in Figure \ref{fig:nicedecomposition} has two children,
%namely $X_{15}$ and $X_5$. $X_{15}$ has terminal $v_3$ below it, whereas $X_5$ has terminal
%$v_1$ below it.

Basically the step is to merge the Steiner tree sets of both children nodes.
Assume the Steiner tree set of both child nodes are $\stset_{X_{c_1} \cup S_{X_{c_1}}}^{|S|}$
and $\stset_{X_{c_2} \cup S_{X_{c_2}}}^{|S|}$ respectively, we shall generate the Steiner tree
set $\stset_{X \cup S_{X_{c_1}} \cup S_{X_{c_2}}}^{|S|}$  at the parent node.

The merge operation is analogously to the vertex insertion traversal we introduced above.
We start from one child node, say $X_{c_1}$. The task is to insert the terminals 
in $S_{X_{c_2}}$ into $\stset_{X_{c_1} \cup S_{X_{c_1}}}^{|S|}$ one by one.
The vertex separator between any terminal in $S_{X_{c_1}}$ and $S_{X_{c_2}}$
is obviously the bag $X$.
The time complexity for the traversal is  $(2tw)^{|S|}$ as well.

To conclude, in the bottom up traversal, at each step the Steiner tree set can be computed
with the time complexity of $(2tw)^{|S|}$. 
Since the number of steps is of $O(h \cdot |S|)$,
%Since the size of a nice tree decomposition $T_G$ is linear to the size of the graph $n$, 
 the overall time complexity 
is $O(h \cdot (2tw)^{|S|})$. This completes the proof. \qed
\end{proof}

\section{Conclusion}

In this paper we presented the algorithm STEIN I, solving the Steiner tree problem over tree decomposed-based index structures.
One requirement for the algorithm is that the sub-Steiner trees in the bags have to be pre-computed. 
%This is a time-consuming task.
However, the computation can be conducted offline. Note that the space for the index storage is $O(tw^{|S|} \cdot |V|)$. 

%The query time is exponential to the number of terminals. Thus the algorithm is not suitable for  problems with large terminal size.
%However, in the keyword search applications, the number of keywords is normally small.
If the number of terminals is considered as a constant, the algorithm is polynomial to the treewidth of the graph $G$,
thus the algorithm can also be applied to the graphs whose treewidth is not bounded, but the relationship $tw \ll |V|$ holds.
As stated in Introduction, finding $tw$ is an intractable problem. Thus even the treewidth of a graph is bounded, 
it is unlikely $tw$ can be obtained from a graph of large size. So in practice we can only find certain width $w$, s. t. $w \ll |V|$,
which can be achieved for the graphs from Web information systems.
There is no obvious correlation between the time complexity and the graph size. In theory, the height of the  tree decomposition ($h$)
is $\log |V|$ for balanced tree and $|V|$ in worst case. However in practice this value is much smaller than the graph size.

\bibliographystyle{plain}
\bibliography{steiner}

\end{document}